\documentclass{article}
\usepackage{frascatiphys,here,amsmath,amssymb,graphicx,subfigure}
\begin{document}
\title{ 
Natural Dirac Neutrinos from Warped Extra Dimension
}
\author{
Jackson M. S. Wu \\
{\em Albert Einstein Center for Fundamental Physics},\\
{\em Institute for Theoretical Physics, University of Bern},\\
{\em Sidlerstrasse 5, 3012 Bern, Switzerland}
}
\maketitle
\baselineskip=11.6pt
\begin{abstract}
Dirac neutrinos arising from gauged discrete symmetry \`a la Krauss-Wilczek are implemented in the minimal custodial Randall-Sundrum model. In the case of a normal hierarchy, all lepton masses and mixing pattern can be naturally reproduced at the TeV scale set by the electroweak constraints, while simultanously satisfy bounds from lepton flavour violation. A nonzero neutrino mixing angle, $\theta_{13}$, is generic in the scenario, as well as the existence of sub-TeV right-handed Kaluza-Klein neutrinos, which may be searched for at the LHC.
\end{abstract}
\baselineskip=14pt
\section{Introduction}
Despite that the nature of neutrino remains unknown, most efforts has been directed at Majorana neutrinos. One common reason cited against the Dirac scenario is that it is hard to make realistic without excessive fine-tuning.

Recently, the Randall-Sundrum (RS) extra-dimensional scenario has become a novel and powerful framework to understand flavour. The crucial feature is that the mass hierarchy of the Standard Model (SM) charged fermions can arise naturally from their ``geography'' in the warped $\mathrm{AdS}_5$ bulk\cite{Geo5D}. The Yukawa couplings need not be fine-tuned, and can be fully ``anarchic'' (i.e. all naturally of order one and patternless). It was seen that the observed quark mass and mixing pattern can be accurately reproduced in this approach\cite{CNWq}, and it is reasonable to expect the same can for Dirac neutrinos with appropriate bulk localisations, which is indeed found to be the case\cite{CNWl}.
\section{The framework}
The setting used is that of the Minimal Custodial RS (MCRS) model\cite{ADMS03}. The bulk gauge group is $SU(2)_L \times SU(2)_R \times U(1)_{B-L}$, which incorporates the custodial symmetry protecting the $\rho$ parameter. It is broken down to the SM at the boundaries by the IR-localised SM Higgs, $H_1$, and by UV boundary conditions (BCs) and scalar vaccum expectation value (VEV). 

Concentrating on the lepton sector, the leptons are embedded as
\begin{align}\label{Eq:leprep}
L_i &=
\begin{pmatrix}
\nu_{iL}\,[+,+] \\
e_{iL}\,[+,+]
\end{pmatrix} \,,\quad
E_i =
\begin{pmatrix}
\tilde{\nu}_{iR}\,[-,+] \\
e_{iR}\,[+,+]
\end{pmatrix} \,, \quad
\nu_{iR}\,[+,+]  \,,
\end{align}
where $i$ is a generation index, $L$, $E$ denote $SU(2)_{L,R}$ doublets respectively, and $\nu_{R}$ the right-handed (RH) neutrino singlet under $SU(2)_{L,R}$. The parity assignment $+$ ($-$) denotes Neumann (Dirichlet) BCs applied to the spinors on the boundary branes. Only fields with the [+,+] parity contain zero-modes that are part of the low energy spectrum of the 4D effective theory after the Kaluza-Klein (KK) reduction.

SM lepton masses arise as usual from Yukawa interactions with the SM Higgs. To generate Dirac masses for the neutrinos, a second Higgs doublet, $H_2$, transforming only under the $SU(2)_L$ is introduced on the IR brane. After electroweak symmetry breaking, the lepton mass matrices in the 4D effective theory take the form
\begin{equation}\label{Eq:RSM}
(M^{RS}_{e,\nu})_{ij} = 
\frac{v_W}{kr_c\pi}
\lambda^{e,\nu}_{5,ij}f^0_{L}(\pi,c^{e,\nu}_{iL})f^0_{R}(\pi,c^{e,\nu}_{jR}) \,, 
\end{equation}
where $kr_c\pi \approx 37\pi$ is the warp factor solving the hierarchy problem, the Higgs VEVs, $v_1 = v_2 = v_W = 174/\sqrt{2}$~GeV, are taken equal for simplicity, $\lambda_5^{e,\nu}$ are the complex dimensionless 5D Yukawa matrice for the charged leptons and neutrinos, $f^0_{L,R}$ are the zero-mode wavefunctions, and $c_{L,R}$ are the localisation parameters (see\cite{CNWl} for more details).
\section{Dirac neutrinos from a gauged discrete $Z_N$}
To have Dirac neutrinos, one needs to forbid Majorana mass terms, which can be easily accomplished via a $U(1)$ symmetry. Now if this $U(1)$ is global, it can be violated by quantum gravity effects. But if it is gauged, it has to be broken to avoid any new massless gauge bosons appearing. However, a gauged discrete $Z_N$ symmetry can survive from a gauged $U(1)$ broken at some very high scale\cite{KW89}, which is enough for Dirac neutrinos, as is well-known.

To implement this in the MCRS model, one extends the bulk gauge group by a $U(1)_X$, and add a scalar UV-localised, $\phi$, charged under it:
\begin{equation}
D_\mu\phi = \left(\partial_\mu - i g_{5X} X_\mu\right)\phi \,,
\end{equation}
where $X_\mu$ is the $U(1)_X$ gauge field, and $g_{5X}$ the coupling constant. The $U(1)_X$ is spontanously broken when $\phi$ acquires a VEV, $v_\phi$. Parametrising the scalar as $\phi = (v_\phi + \rho)e^{i\eta/v_\phi}$, one sees that the Goldstone field, $\eta$, can be removed by a gauge transformation and a concomitant fermion field redefinition:
\begin{equation}
X_\mu \rightarrow X_\mu - \frac{1}{g_{5X}}\frac{\partial_\mu\eta}{v_\phi} \,, \qquad
f \rightarrow f\exp\left(i\frac{\eta}{v_\phi}Q_X\right) \,.
\end{equation}
The $Z_N$ symmetry then emerges if $Q_X$, the fermion charge under the $U(1)_X$, is rational but nonintegral.
\section{Viability}
To see if the Dirac neutrinos thus implemented can be realistic and natural, a parameter space scan is performed searching for configurations that give rise to the observed lepton mass (at a running scale of 1 TeV) and the bi-large mixing pattern while satisfying the current lepton flavour bounds at the same time. 

For the search, the KK scale is set to 3 TeV to satisfy electroweak precision test constraints. The 5D Yukawas are taken to be $|\lambda_{5,ij}| \in [0.5,2.0]$ for perturbativity. It is found that viable configurations compatible with all current data and bounds arise only for the case of neutrino the normal hierarchy. Displayed in Table~\ref{tb:clep} are the range of lepton localisation parameters for the viable configurations. In Table~\ref{tb:nup}, the resulting range of neutrino mass and mixing parameters are listed. All viable configurations have charged lepton masses $\{m_e,\,m_\mu,\,m_\tau\} = \{0.496,\,105,\,1780\}$~MeV. Note that $\theta_{13}$ is generically nonzero in all the viable configurations found.
\begin{table}[ht]
\centering
\begin{tabular}{c|c}
Lepton species & Parameter range \\
\hline
$\{c_{L_1},\,c_{L_2}\,c_{L_3}\}$ & \{(0.583, 0.588), (0.533, 0.548), (0.500, 0.502)\} \\
$\{c_{E_1},\,c_{E_2}\,c_{E_3}\}$ & \{-0.728, -0.721, (-0.601, -0.588), (-0.520, -0.523)\} \\
$\{c_{\nu_{R1}},\,c_{\nu_{R2}}\,c_{\nu_{R3}}\}$ & \{(-1.33, -1.22), (-1.36, -1.22), (-1.38, -1.22)\}
\end{tabular}
\caption{\label{tb:clep} Range of lepton localisation parameters for the viable configurations.}
\end{table}

\begin{table}[ht]
\centering
\begin{tabular}{c|c}
Lepton parameters & Paramter range \\
\hline
$\{m_{\nu_1},\,m_{\nu_2}\,m_{\nu_3}\}$ (meV) & \{(0.096, 1.4), (8.5, 9.1), (47, 53)\} \\
$\{\theta_{12},\,\theta_{23},\,\theta_{13}\}$ ($^\circ$) & \{(32, 39), (36, 53), (1.9, 12)\}\\
$\delta_{CP}$ (rad.) & $[0,\,2\pi)$
\end{tabular}
\caption{\label{tb:nup} Range of neutrino mass and mixing parameters from the viable configurations.}
\end{table}
\section{Phenomenology}
\subsection{Light KK neutrinos}
KK excitations are characteristic in the extra-dimensional scenarios. But with the KK scale at 3~TeV, they are already hard to produce and thus detect at the LHC. However, depending on their localisations, the RH $(-+)$ KK neutrinos, $\tilde{\nu}_{iR}$, can be much lighter in comparison. For the viable range of $c_E$, one has for $\tilde{\nu}_{iR}$ the first KK masses (determined from their BCs\cite{CNWl})
\begin{equation}
m_{\tilde{\nu}_1}: 175 - 222\,\mathrm{MeV} \,,\quad
m_{\tilde{\nu}_2}: 16 - 24\,\mathrm{GeV} \,,\quad
m_{\tilde{\nu}_3}: 168 - 180\,\mathrm{GeV} \,.
\end{equation}
\subsection{Low energy constraints}
The RH $(-+)$ KK neutrinos couple to the SM W and Z primarily through gauge mode mixings arising from SM Higgs interactions on the IR brane. These couplings can be parametrised as
\begin{equation}
W\tilde{\nu}_{iR}e_{iR}: \frac{g_L}{\sqrt{2}}\,r_i(c_{E_i}) \,, \quad
Z\bar{\tilde{\nu}}_{iR}\tilde{\nu}_{iR}:
\frac{g_L}{\cos\theta_W}\gamma^\mu\!\left[z_{Li}(c_{E_i})\hat{L} + z_{Ri}(c_{E_i})\hat{R}\right] \,,
\end{equation}
where $g_L \equiv e/\sin\theta_W$, $\theta_W$ the Weinberg angle, and $\hat{L}$, $\hat{R}$ the usual chiral projectors. The quantities $r_i$, $z_{Li}$ and $z_{Ri}$ involve gauge couplings and products of wavefunction overlap integrals. For simplicity $g_L = g_R$ is assumed.

With $m_{\tilde{\nu}_1} > m_\pi$, the charged kaon decay $K^+ \rightarrow e^+\tilde{\nu}_1$ sets the most stringent limit with the bound $|r_1|^2 < 10^{-6}$ for a 160 to 220 MeV neutrino~\cite{PDG08}. Another limit comes from the LEP invisible Z decay measurement, which gives $N_\nu=\Gamma_{inv}/\Gamma_\nu^{SM} = 2.9840 \pm 0.0082$\cite{ALEPH06}. Since only $\tilde{\nu}_1$ is light enough to escape the detector without leaving tracks, this sets a bound on $\Gamma(Z\rightarrow\bar{\tilde{\nu}}_1\tilde{\nu}_1)$ and thus
$z^2_{L1} + z^2_{R1} \leq 0.096 \; (95\%\,\mathrm{CL})$. For the viable range of $c_E$, these constraints turns out to be very weak, as $ r_1\sim O(10^{-6})$, while $z_{L1} \sim z_{R1} \sim O(10^{-2})$.
\subsection{Decay modes and LHC production}
The decay modes of the RH $(-+)$ KK neutrinos depends crucially on their masses. For the heaviest $O(100)$~GeV $\tilde{\nu}_3$, a decay predominantly into $\tau\,W$ is expected with a width
$\Gamma_{\tilde{\nu}_3} \sim 1.5 \times 10^{-6}$~GeV for the viable range of $c_E$.

For the (much) lighter $\tilde{\nu}_{1,2}$, three body decays are dominant. For $\tilde{\nu}_1$, the dominant decay channel is the charged current (CC) decay $\tilde{\nu}_1 \rightarrow e e^+\nu_e$~\footnote{The $\tilde{\nu}_1 \rightarrow e\mu^+\nu_\mu$ contribution is phase space suppressed, the $e\,\pi$ mode is negligible, while the virtual $Z$ mediated amplitudes are unimportant.} with a width 
$\Gamma^\mathrm{CC}_1 \sim 0.73 \times 10^{-17}|r_{1}|^2$~GeV, corresponding to a lifetime of $\tau_{\tilde{\nu}_1} \sim 2.3 \times 10^{4}$~s for the viable range of $c_E$. For $\tilde{\nu}_2$, the main decays channels are $\tilde{\nu}_2\rightarrow\mu\bar{l}\nu_l,\,\mu\bar{u}d,\,\mu\bar{c}s$. The total width is estimated to be $\Gamma_2^\mathrm{CC} \sim 0.027|r_2|^2$~GeV, and so a lifetime 
$\tau_{\tilde{\nu}_2} \sim 1.2 \times 10^{-15}$~s for the viable range of $c_E$.

At the LHC, due to the large background $\tilde{\nu}_1$ is not expected to be seen given that
$m_{\tilde{\nu}_1} \ll 1$~GeV. For $\tilde{\nu}_2$, it can be detected via the process $u\bar{d}\rightarrow\tilde{\nu}_2\mu^+\rightarrow\mu^+\mu^-e(\tau)\bar{\nu}$. The final state will involve apparent lepton flavor violation plus missing energy with the $\mu^+\mu^-$ pair not in resonance, characteristic of heavy neutrino signatures. Similarly, $\tilde{\nu_3}$ can be detected via the process $u\bar{d}\rightarrow\tau^+\tilde{\nu}_3\rightarrow\tau^+\tau^-W$, where a $W$ jet plus $\tau$ jets are expected with the $\tau$ jets not in resonance. For the viable range of $c_E$, the total production cross section at $\sqrt{s} = 14$~TeV is estimated to be $\sim 0.3$~fb and $\sim 10^{-3}$~fb for $\tilde{\nu}_{2,3}$ respectively.
\section{Conclusion}
It is shown that Dirac neutrinos can be naturally implemented in the MCRS setting with a
$SU(2)_L\times SU(2)_R \times U(1)_{B-L}\times U(1)_X$ bulk gauge symmetry group \`a la Krauss-Wilczek. For normal neutrino hierarchy only, lepton masses and mixing patterns can be successfully reproduced with just the RS anarchic 5D flavour structure, at the TeV scale, and at the same time still satisfy the lepton flavour bounds. 

In the Dirac neutrino scenario presented, $\theta_{13}$ is generially neither zero nor small, which can be tested at the upcoming long-baseline experiments. There are also light KK neutrinos in the spectrum, of which the $O(200)$~MeV $\tilde{\nu}_1$ is too light to be picked out at the LHC but has a long life time of $O(10^4)$~s, the heavy $O(200)$~GeV $\tilde{\nu}_3$ has to small a production rate, leaving only the $O(20)$~GeV $\tilde{\nu}_2$ possible to be searched for at the LHC.
\section{Acknowledgements}
JW thank LNF for the hospitality and support provided during the 2nd Young Researchers Workshop.
\end{document}